\documentclass[twocolumn]{aastex62}

\graphicspath{{./}{figures/}}
\usepackage{gensymb}

\accepted{to the Astronomical Journal}

\shorttitle{EDD: The CMDs/TRGB Catalog}
\shortauthors{Anand, Rizzi, Tully et al.}

\begin{document}

\title{\textbf{The Extragalactic Distance Database: The Color-Magnitude Diagrams and \\ Tip of the Red Giant Branch Distances Catalog}}

\correspondingauthor{Gagandeep S. Anand}
\email{gsanand@hawaii.edu}

\author{Gagandeep S. Anand}
\affiliation{Institute for Astronomy, University of Hawaii, 2680 Woodlawn Drive, Honolulu, HI 96822, USA}

\author{Luca Rizzi}
\affiliation{W.M. Keck Observatory, 65-1120 Mamalahoa Highway, Kamuela, HI 96743, USA}

\author{R. Brent Tully}
\affiliation{Institute for Astronomy, University of Hawaii, 2680 Woodlawn Drive, Honolulu, HI 96822, USA}

\author{Edward J. Shaya}
\affiliation{Astronomy Department, University of Maryland, College Park, MD 20743, USA}

\author{Igor D. Karachentsev}
\affiliation{Special Astrophysical Observatory, Russian Academy of Sciences, Nizhniy Arkhyz, Karachai-Cherkessia 369167, Russia}

\author{Dmitry I. Makarov}
\affiliation{Special Astrophysical Observatory, Russian Academy of Sciences, Nizhniy Arkhyz, Karachai-Cherkessia 369167, Russia}

\author{Lidia Makarova}
\affiliation{Special Astrophysical Observatory, Russian Academy of Sciences, Nizhniy Arkhyz, Karachai-Cherkessia 369167, Russia}

\author{Po-Feng Wu}
\affiliation{National Astronomical Observatory of Japan, Osawa 2-21-1, Mitaka, Tokyo 181-8588, Japan}

\author{Andrew E. Dolphin}
\affiliation{Raytheon, 1151 E. Hermans Road, Tucson, AZ 85706, USA}

\author{Ehsan Kourkchi}
\affiliation{Institute for Astronomy, University of Hawaii, 2680 Woodlawn Drive, Honolulu, HI 96822, USA}

\begin{abstract}
The Extragalactic Distance Database (EDD) was created as a repository for high quality, redshift-independent distances. A key component of EDD is the Color Magnitude Diagrams/Tip of the Red Giant Branch (CMDs/TRGB) catalog, which provides information on the stellar content of nearby galaxies observed with the Hubble Space Telescope (\textit{HST}). Here we provide a decadal update to this catalog, which has now doubled in size to over 500 galaxies. We highlight the additions to our data reduction and analysis techniques, and provide examples of the science that has been made possible with this large data set. We find the TRGB to be a reliable measure for distance, and we aim to extend its distance coverage with \textit{HST} to every galaxy within 10~Mpc. In the near-future, the combination of the James Webb Space Telescope and the Nancy Grace Roman Space Telescope will dramatically increase the number of targets within our grasp.

\end{abstract}

\section{Introduction} \label{sec:intro}

\subsection{Background}

The Hubble Space Telescope (\textit{HST}) has revolutionized our understanding of the stellar content of nearby galaxies. Deep color-magnitude diagrams (CMDs) can be extracted from relatively modest exposure times, allowing us to easily obtain information that would be very difficult to derive from ground based observations. A key feature on these CMDs is the tip of the red giant branch (TRGB), a stage of stellar evolution that can be used to determine distances with accuracies rivaling that of Cepheid variables \citep{1993ApJ...417..553L,1996ApJ...461..713S,2007ApJ...661..815R,2016AJ....152...50T, 2017AJ....154...51M,2018ApJ...858...62K,2019ApJ...882...34F,2021MNRAS.501.3621A}.

Low mass ($\lessapprox 2 M_{\odot}$) main-sequence stars will, later in their lives, track all the way up the red giant branch. At the end of this phase, their degenerate helium cores will become hot enough to begin to fuse helium in a runaway process known as the helium flash. This short lived event is followed by a transition to the horizontal branch, a phase at which these stars become much dimmer than their red giant branch precursors. Above the RGB lie asymptotic giant branch stars, which are highly evolved low and intermediate mass stars. The rapid transition of low mass stars from the RGB to the horizontal branch leaves a distinct observational signature in the form of a discontinuity in the luminosity function of stars at the magnitude of the TRGB. Given that the maximum mass of the helium core at the point of the helium flash is constant, we can use the magnitude of the TRGB as a standard candle. Modest corrections for metallicity are required, mostly due to the effects of line-blanketing.

The TRGB has several key advantages over Cepheid variables, which are the main alternative standard candle for nearby galaxies. First, essentially every nearby galaxy has an old stellar population, and thus red giant branch stars. The same can not be said for Cepheid variables, whose presence requires the existence of a relatively young ($< \sim$100 Myr) stellar population. There are many nearby galaxies (ellipticals, dwarf spheroidals, etc.) that do not have a significant number of Cepheid variable stars. 

Another key advantage of the TRGB is the efficiency with which the requisite observations can be acquired. Precise and accurate distances ($\sim$5$\%$) to unobscured galaxies out to 10 Mpc can be obtained with just one orbit with \textit{HST} \citep{2015ApJ...805..144K,2018MNRAS.474.3221M,2019ApJ...872...80C}. Heavily obscured galaxies can be observed in the near-infrared, where the impact of extinction from foreground dust is minimized \citep{2017ApJ...835...78R, 2019ApJ...872L...4A, 2019ApJ...880...52A}. There is no need for multiple epochs of observations to create a light curve, as is needed to identify Cepheids. The method has even been applied to targets in the $\sim$20-30~Mpc range, though such feats require significant observing time \citep{2008AJ....136.1482S,2017ApJ...836...74J}.

\subsection{Motivation for the Catalog}

Given the ability to obtain precise distances to nearby galaxies at relatively low costs, the TRGB has been widely used by the larger community. However, the applications in the literature are not uniform, as they have been performed by many different authors. Differences include the photometry packages used, the manner of determining the magnitude of the TRGB, and the calibration used for determining the absolute magnitude of the TRGB, among others. 

The Cosmicflows program \citep{2008ApJ...676..184T,2013AJ....146...86T,2016AJ....152...50T} aims to obtain high-quality redshift-independent distances, which are then used in conjunction with velocity information to determine peculiar velocities, i.e. the motions of galaxies after accounting for their velocities from the Hubble flow. Analysis of this data has led to insights into the large-scale structure and dynamics of the Universe out to several hundred megaparsecs \citep{2014Natur.513...71T, 2017NatAs...1E..36H, 2017ApJ...847L...6C, 2017ApJ...850..207S, 2019ApJ...880...24T, 2019MNRAS.489L...1D, 2020ApJ...897..133P}. The Cosmicflows data, along with other relevant catalogs, are all publically available at the Extragalactic Distance Database\footnote{\url{https://edd.ifa.hawaii.edu}} \citep{2009AJ....138..323T}.

Cosmicflows, now working towards its fourth iteration with distances to $\sim$50,000 galaxies, uses Cepheid and TRGB distances not only for the determination of nearby large-scale structure, but also as calibrators for other distance methods (e.g. the Tully-Fisher relation, \citealt{2020ApJ...896....3K}). Small scale systematic differences in the applications of the TRGB method in the literature would propagate into the other distances obtained from Cosmicflows. Given the goals of the project, this is not acceptable. 

To reduce competing systematics and ensure a common scale for the TRGB distances used in Cosmicflows, we undertook to perform TRGB measurements for every galaxy for which the requisite data was available from \textit{HST}. The details of the initial public release of this catalog were presented by \cite{2009AJ....138..332J}, with an initial release of $\sim$250 galaxies. The aim of this paper is to provide a decadal update upon that work. We expand upon the descriptions provided in that work, and include updates to our previously established procedures. We include highlights of science cases made possible with our catalog, and conclude with a brief outlook on the future of the TRGB.


\section{Data}

\begin{figure*}
\centering
\epsscale{1.2}
\plotone{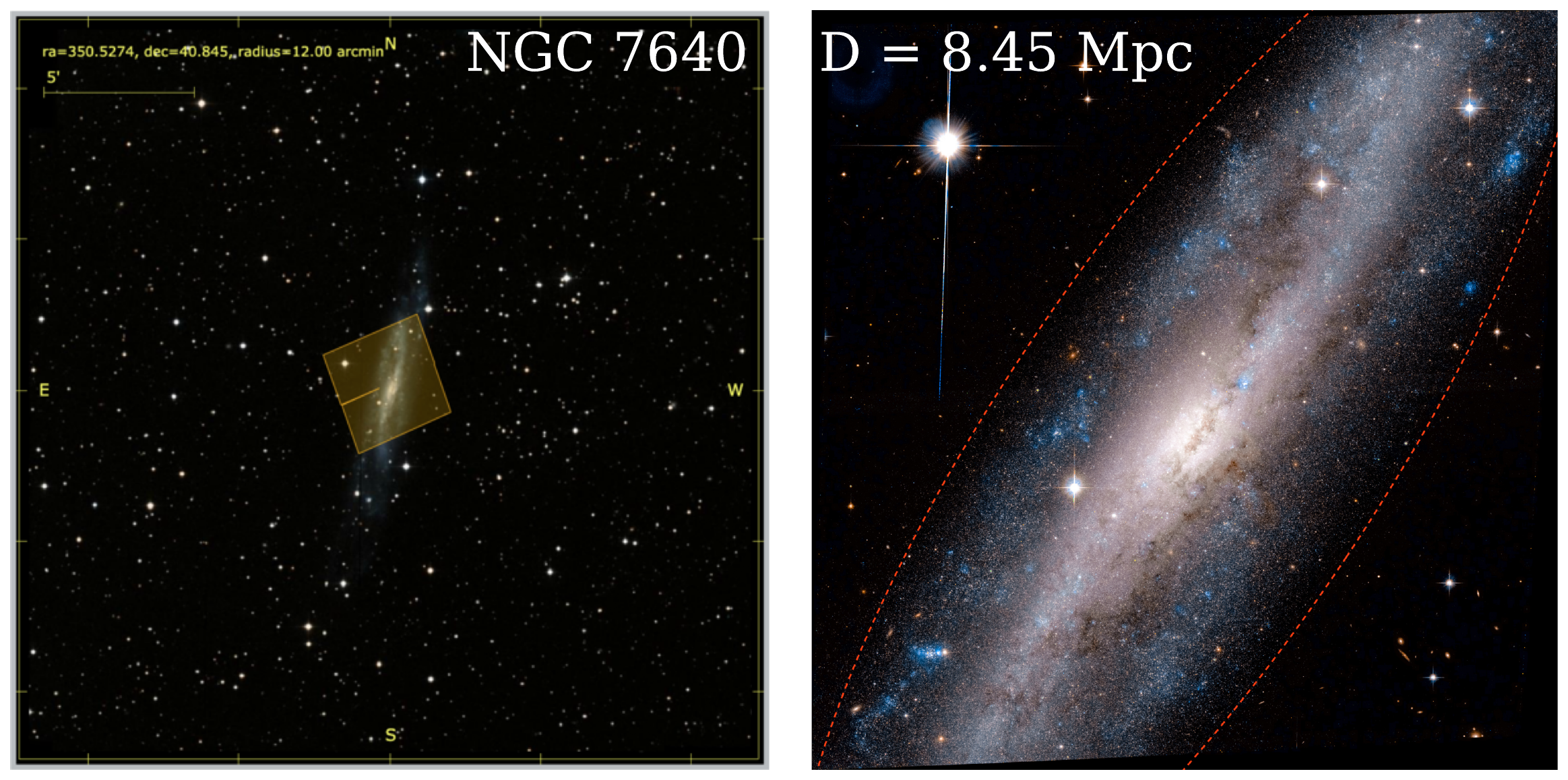}
\caption{\textbf{Left)} Footprint of the ACS/WFC imaging of NGC 7640 overlayed on DSS imaging (obtained from the Hubble Legacy Archive). \textbf{Right)} Color image of the galaxy generated from the F606W+F814W imaging. Only stars exterior to the red ellipse were used for the final determination of the TRGB. A higher resolution of this color image is available on EDD.}
\label{NGC7640colors}
\end{figure*}

\subsection{Imaging}

The largest component of the data in our catalog has been taken with the Advanced Camera for Surveys/Wide Field Channel (ACS/WFC), which images a 202$''$ $\times$ 202$''$ region with 0.05$''$ pixel scale. Wide Field Camera 3 (WFC3) also has UV/optical capabilities, with a sharper pixel scale of 0.04$''$ but a field of view of 160$''$ $\times$ 160$''$ (63$\%$ of the sky coverage of ACS/WFC) and slightly reduced sensitivity toward the red. ACS/WFC, with its larger field of view and higher sensitivity in the F606W and F814W bands, is our camera of choice given that our focus is on red giant stars located in regions of galaxies with minimal crowding. With the F606W and F814W bands, the CMD is constructed as a F814W vs. F606-F814W diagram. F814W is the preferred filter to measure the \emph{apparent} TRGB, as the \emph{absolute} magnitude of TRGB stars is relatively consistent in this filter over a large range of observed colors. The secondary filter is used to isolate the red giant branch from stars of similar magnitudes but different colors (e.g. upper main-sequence stars). For a more in-depth discussion of issues regarding the zero-point and color calibration of the TRGB, see \S 4.4. While F606W is the preferred secondary band to measure colors, observations in F555W and even F475W may be used. The F606W band is preferred because the more blueward filters can result in loss of parts of the RGB from the observed CMD, limiting their usefulness especially in cases of increased metallicities. 

The WFC3/IR detector provides us with infrared coverage with our choice of F110W and F160W filters (approximately J $\&$ H bands). The field of the WFC3/IR detector is 123$''$ $\times$ 136$''$, with a pixel scale of  0.13$''$. Observations with WFC3/IR are particularly useful in cases where there is a large amount of foreground extinction. Unlike the case with the optical filters, we perform the TRGB measurement with \emph{both} F110W and F160W bands. This is because of a trade-off that exists between reddening and the absolute magnitude of the TRGB in the near-infrared. Unlike the F814W filter, the absolute magnitude of the TRGB is a rather sensitive function of metallicity in the near-infrared bands. Foreground extinction is most minimized in the F160W band (decreased by $\sim$40$\%$ compared to F110W), however this comes at the cost of an increased metallicity dependence (by a factor of $\sim$1.5) of the TRGB. Performing the TRGB measurement in both F110W and F160W provides us with two separate (but not completely independent) measurements to perform a cross-check of our results.

Lastly, while the Wide Field and Planetary Camera~2 (WFPC2) has been retired, early observations made with WFPC2 represent a significant portion of the data in our catalog. There are 133 galaxies with only WFPC2 imaging, while supplemental WFPC2 imaging is available for many of the ACS or WFC3 targeted galaxies.

\subsection{Footprints}
With larger galaxies, the \textit{HST} field of view typically only covers a portion of the galaxy. For most dwarf galaxies, the \textit{HST} field of view covers the entire target. In either case, knowledge of the larger context may aid our end users in selection of data for further work. To place our observations into the proper environmental context, we provide footprints of the data.  

We initially obtained these footprints from the Hubble Legacy Archive (HLA)\footnote{\url{https://hla.stsci.edu/}}. The \textit{HST} instrument footprints from the HLA are displayed over color images from the Digitized Sky Survey (DSS), with a default size of 24$'\times$24$'$. A sample footprint of an ACS/WFC observation of NGC~7640 is shown in the left panel of Figure \ref{NGC7640colors}. The HST field of view covers the majority of the main disk of the galaxy. The data for this target was obtained as part of GO-13442 (PI R. Brent Tully). As of January 2021, the HLA interface no longer provides footprints, and we now retrieve them directly from the Mikulski Archive for Space Telescopes (MAST)\footnote{\url{mast.stsci.edu}}. 

\subsection{Color Images}

\begin{figure*}[t]
\epsscale{1.2}
\plotone{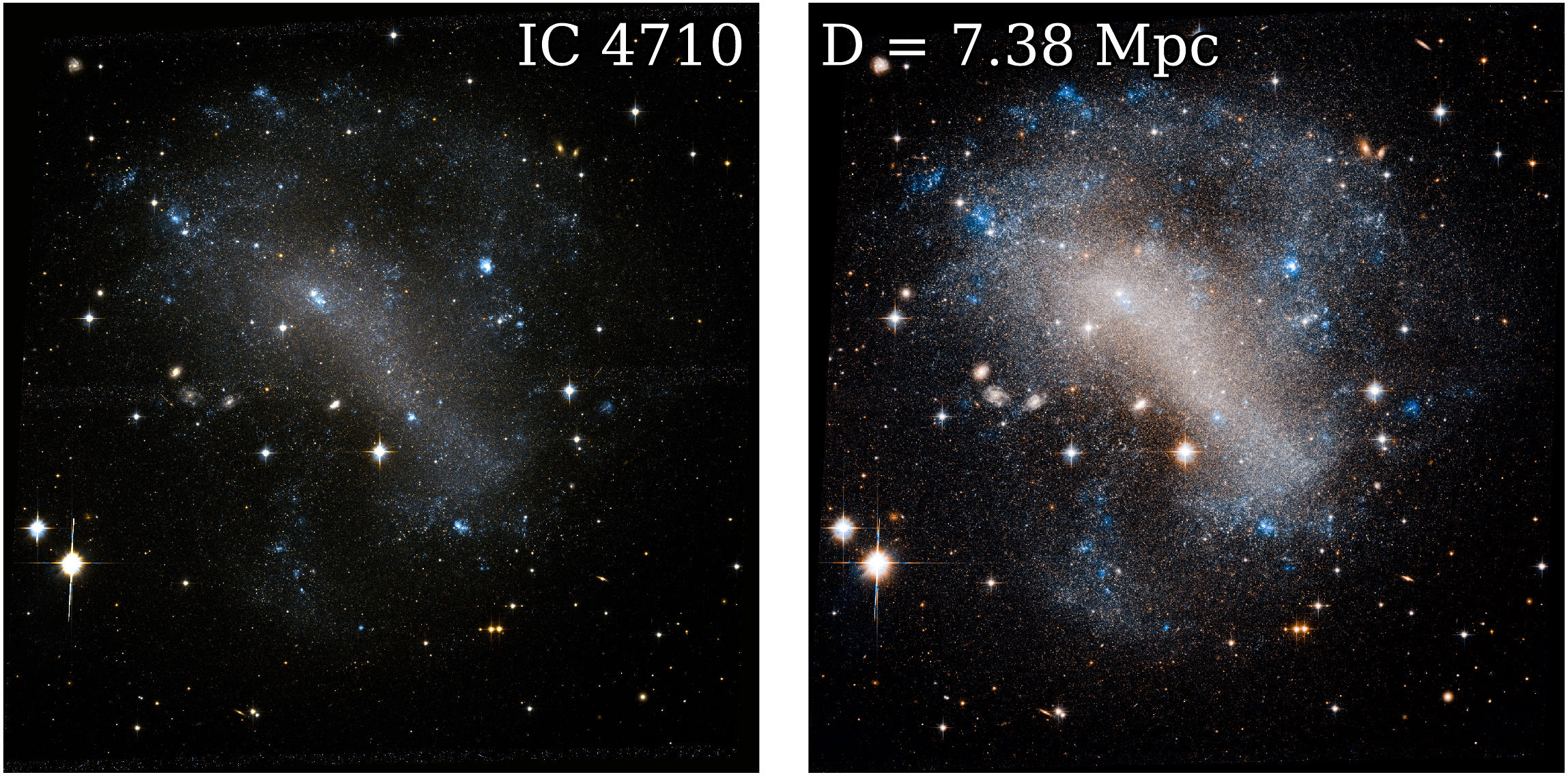}
\caption{A comparison between the catalog's old and new optical color images for IC 4710. A higher resolution version of the new image is available on EDD.}
\label{IC4710}
\end{figure*}

In addition to providing footprints, we also use our two-filter data to create pseudo-color images of each field. These color images help to highlight the diversity of stellar environments present in each image, and allow the end user to examine our catalog for regions of interest (e.g. star clusters). Color images also serve an important public outreach purpose. Data from several targets in our programs have been used by us and others in the community to create images for the Astronomy Picture of the Day\footnote{\url{https://apod.nasa.gov/}} or the ESA/Hubble Astronomy Picture of the Week\footnote{\url{https://www.spacetelescope.org/images/potw/}}.  

We use DrizzlePac \citep{2015ASPC..495..281A} to combine the individual flc images into a deep, drizzled image for each filter. Since we only have photometry in two filters, we use an average of the two images for the green image. The three individual color images are then combined with APLpy \citep{2012ascl.soft08017R} $\&$ Montage \citep{2010ascl.soft10036J, 2017ASPC..512...81B}. The images are then exported to Lightroom where adjustments to the color-stretches are made, providing the final color images with a high dynamic range. Our focus is on separating the young and old stellar populations; we do not give rigorous attention to creating a ``true-color" image.

A sample color image for NGC~7640 is shown in the right panel of Figure \ref{NGC7640colors}. The HST/ACS field of view covers the majority of the disk, and thus highlights the key regions of this spiral galaxy. Prominently visible are the galactic bulge, dust lanes, numerous blue star clusters, as well as the many red giant branch stars. This overall procedure for developing our color images is relatively new, and thus many of the color images were developed with older techniques. We are in the process of updating all of these images. Figure \ref{IC4710} shows a comparison between the old and new F606W+F814W color images for IC4710. The updated image does a significantly better job at displaying both the high and low surface-brighness regions in this dwarf irregular galaxy.

\begin{figure*}[t]
\epsscale{1.2}
\plotone{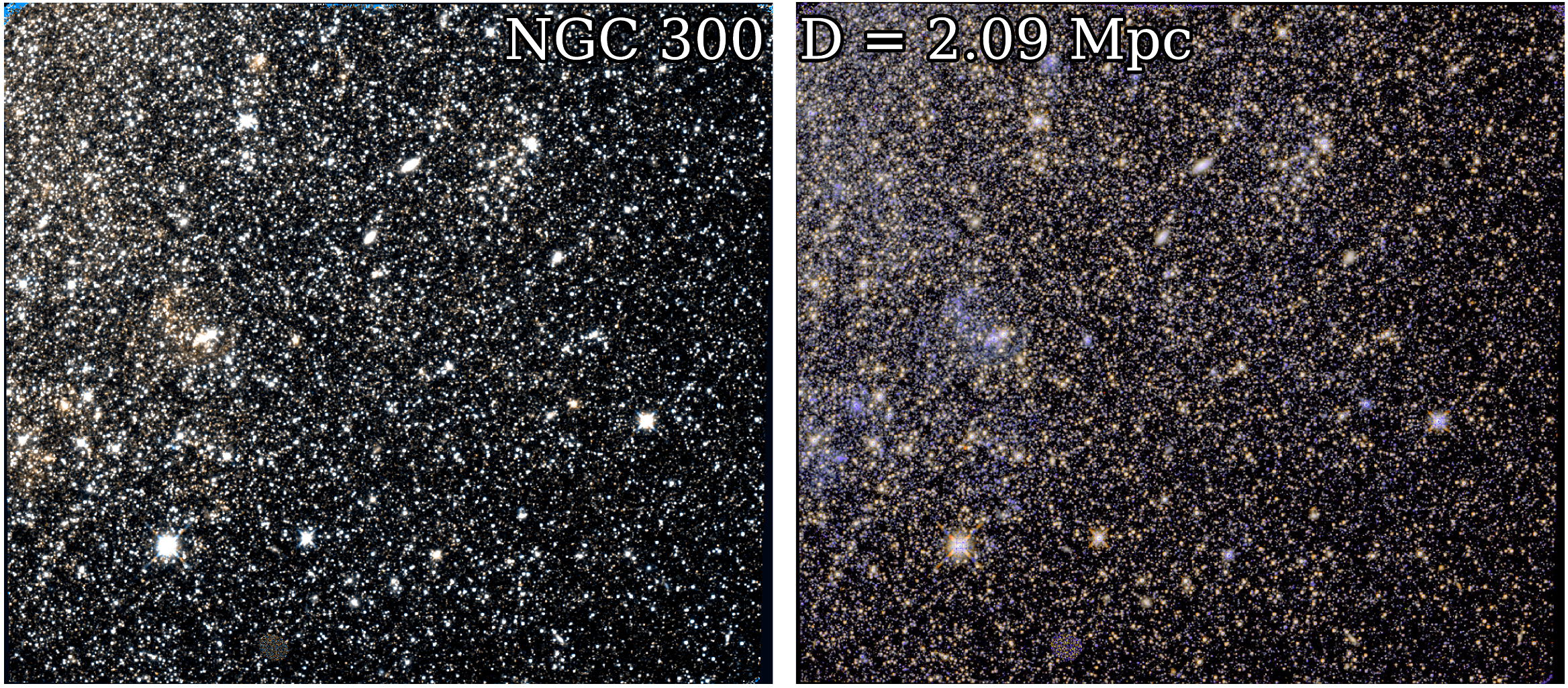}
\caption{A comparison between the catalog's old and new near-infrared color images for a portion of NGC~300. A higher resolution version of the new image is available on EDD.}
\label{NGC300}
\end{figure*}

With our original near-infrared studies, the applied color stretches tended to make the images appear essentially monochromatic. We have rebuilt these near-infrared images to appropriately separate the distinct stellar populations. Figure \ref{NGC300} shows our previous and current color images for NGC~300, taken as part of SNAP-11719 (PI J. Dalcanton, \cite{2012ApJS..198....6D}). The new color image gives reasonable definition to the star-forming nebulae, as well as to the blue upper-main sequence and red giant stars.

\section{Photometry}

\subsection{Data Reduction}

One of the two key science products of this catalog is the information on the resolved stellar populations in nearby galaxies. Accurate photometry is key to our main science goals. Aperture photometry is not reliable due to the levels of crowding.  The point-spread function (PSF) of the instruments on \textit{HST} is carefully monitored and well-understood, motivating our use of PSF photometry. 

We perform PSF photometry using the latest versions of the DOLPHOT \citep{2000PASP..112.1383D, 2016ascl.soft08013D} package. Some words are in order about DOLPHOT. We obtain individual images from MAST in the \textit{flc} format, which are calibrated through the STScI pipeline and CTE-corrected. In general, photometry recovered from DOLPHOT has been shown to be consistent with that from other PSF photometry packages, such as DAOPHOT \citep{2017ApJ...835...28J}. 

DOLPHOT requires the user to select a single image for use as a reference frame in aligning all other images. For a typical case with two-filter observations taken in a single \textit{HST} orbit, MAST typically provides a single drizzled image for each filter set. Either of these can then be used as the reference image. We note that no photometry is performed on this \textit{drc} frame- it is only used for alignment purposes (drizzled images result in sub-optimal PSF photometry due to resampling).

In situations with observations occurring over several orbits or visits, there likely are several drizzled images (one for each orbit or visit).  In these cases, we create a deep, drizzled image (\textit{drc}) using the AstroDrizzle task in DrizzlePac \citep{2015ASPC..495..281A}. We use all \textit{flc} frames in one of the available filters (with preference given to one with deeper observations). In cases with separate visits, the relative \textit{HST} astrometry is often displaced by as many as $\sim$15 pixels. When using DOLPHOT with ``UseWCS = 1" (to use the WCS header information for alignment), this offset is too large for DOLPHOT to find the correct alignment solution. We correct for this by using the procedure described in Section 5 of \cite{2015AJ....149...51C}. In short, we use \textit{tweakreg} \citep{2015ASPC..495..281A} to correct the relative astrometry between the individual images before creating the drizzled reference frame. We then use the individual \textit{flc} frames that have been updated with the new \textit{WCS} information when performing our photometry.

DOLPHOT also requires a file that specifies photometry parameters (e.g. number of images, aperture radius, etc.). We use the recommended parameters based on the specific instrument and camera combinations. These parameters are available in the DOLPHOT user's manuals\footnote{\url{http://americano.dolphinsim.com/dolphot/}}.

In addition to the  photometry, we also perform artificial star experiments with DOLPHOT. Based on the positions and magnitudes of the observed photometry, we inject $\sim$100,000 artificial stars into the images and perform photometry with the same parameters as done for the real stars. The results from these experiments are then used to help determine the magnitude of the TRGB and quantify the associated errors.

The raw output of the DOLPHOT photometry is a file ending in \textit{*.phot}. This file contains information on the photometry for the individual frames, as well as that for the combined observations. This raw photometry file contains many marginal sources that are not well-resolved stars, and needs to be filtered before use. The DOLPHOT manual recommends several parameters useful as filters of the output photometry. These include sharpness of the point-source, object type, and error flags (e.g. possible saturation, star not fully on chip, etc.), among others. Our selection criteria filter for stars with a type = 1 or 2 (good or faint stars), and no error flags. We also filter our photometry list based on crowding, sharpness, and signal-to-noise. Many studies have been carried out to determine the optimal recovery of high-quality point sources with varying DOLPHOT selection criteria. At present, for optical photometry we filter the \textit{*.phot} file based on criteria similar to those described in the studies by \cite{2009ApJS..183...67D}, \cite{2011ApJS..195...18R}, and \cite{2017AJ....154...51M}, with flexibility for the details of the specific observation (exposure times, field-crowding level, etc.). For near-infrared photometry, we typically use criteria described by \cite{2012ApJS..198....6D}. 

The culling of the photometry results in a new file, which we denote with the \textit{*.phot2} suffix. This file no longer contains the photometry information for the individual frames, but just the globally determined photometry. In cases with highly crowded environments or many individual frames, a single raw photometry file can exceed many gigabytes. We typically only supply the \textit{*.phot2} photometry file through our website.

\begin{figure*}
\centering
\epsscale{1.2}
\plotone{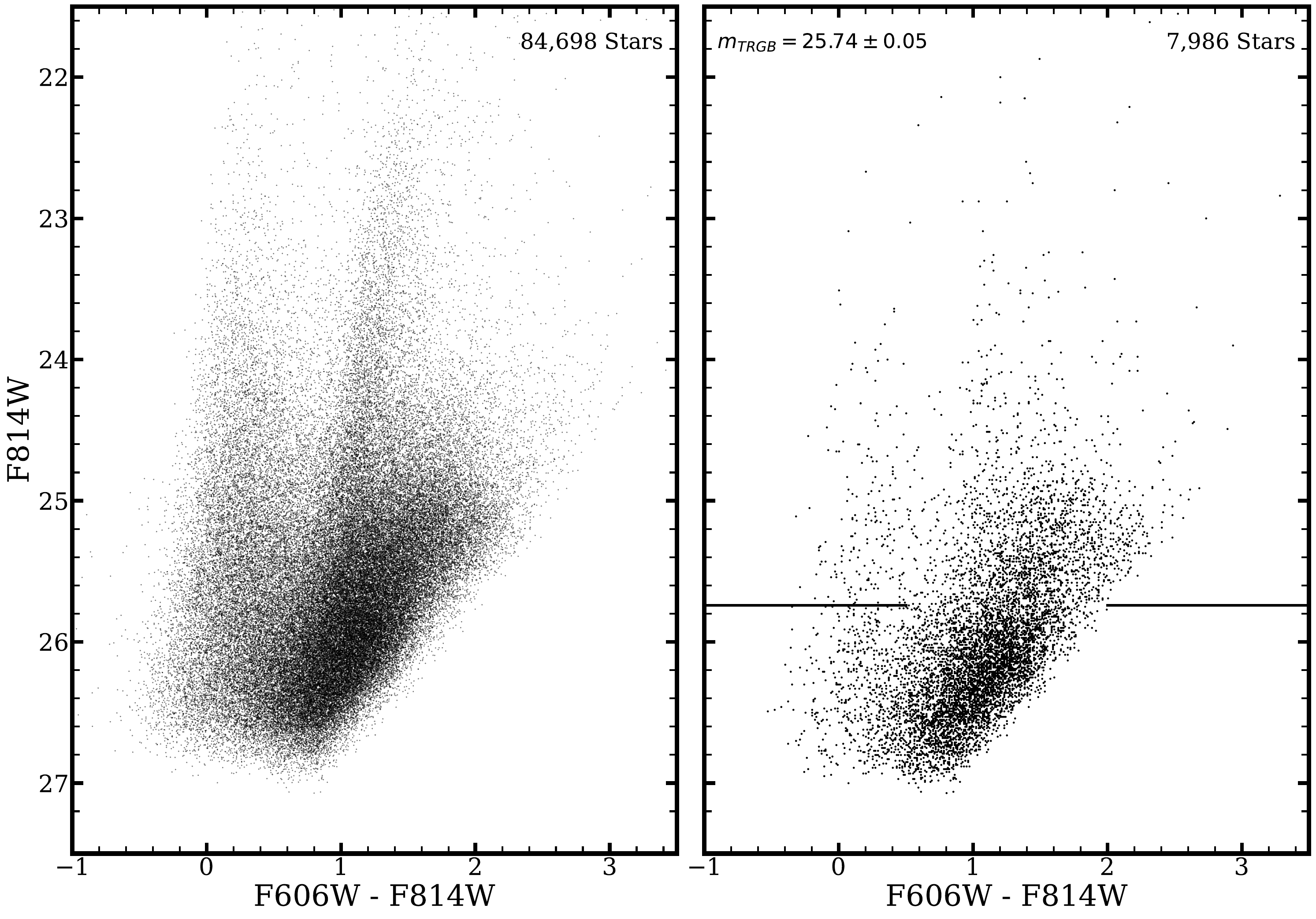}
\caption{\textbf{Left)} CMD of the full field of NGC~7640 (see Figure 1) after performing quality cuts on the photometry. \textbf{Right)} A CMD restricted to the halo (see the right-panel of Figure \ref{NGC7640colors}). The magnitude of the TRGB is marked with the black line, with the gap marking the color region used for fitting.}
\label{NGC7640CMDS}
\end{figure*}

\subsection{Color Magnitude Diagrams}

In all cases, we display a color-magnitude diagram of the full field of view as the first image under each proposal ID. Our TRGB measurement is shown by a solid black line, where the break signifies the color range of the CMD used for the determination of the TRGB.  The \textit{*.phot2} photometry file is available for display/download by left/right clicking on this CMD.  We also provide an additional ``Clipped CMD" in cases where we limited our TRGB analysis to a specific region of the field of view. In these cases, we only show the measurement line on this second, restricted CMD. Examples of both the full-field CMD and a clipped CMD are shown in Figure \ref{NGC7640CMDS}. We do not correct for foreground extinction in any of the displayed color-magnitude diagrams. These corrections are only applied after we determine the TRGB magnitude.

\section{Tip of the Red Giant Branch Determination}

\subsection{Field Selection}

As alluded to previously, there are occasions when it is best to limit our analysis to a subset of the field of view. First take the example of a dwarf spheroidal galaxy that only covers a small fraction of \textit{HST's} field of view. In this situation, we can easily isolate the galaxy from the background, and limit contamination from background point sources (e.g., galactic stars, unresolved galaxies). We usually perform such a procedure either with a simple Python script, or by using \textit{glue}\footnote{\url{http://glueviz.org/}}, a Python package designed for easy data exploration, visualization, and manipulation. \textit{glue} allows us to link together the photometry file and the reference image, which allows for seamlessly restricting photometry to only particular regions of an image.

If the target is instead a small dwarf irregular galaxy with a few prominent star forming regions, we will typically follow a similar procedure, with the added step of blocking out the star-forming regions from our final CMD. In this way, we can limit contamination from stellar populations that are not accounted for in our simple luminosity function model. These contaminants include AGB stars that are reddened by internal dust to the point which they may appear as RGB stars, as well as red supergiants, whose ``plume" feature on a CMD lies near the RGB sequence. 

In cases like that of NGC~7640 (see Figure \ref{NGC7640colors}), the most effective way to limit contamination is to restrict the CMD to the halo of the galaxy. Main sequence stars will be present throughout the entire main disk of the galaxy, but only red giants will populate the halo. In cases like NGC~7640 (see the dashed ellipse plotted in Figure \ref{NGC7640colors}), we can simply remove regions with young stars (i.e. the entire visible disk) that lie in the inner region of the galaxy with a tool like \textit{glue}. For more complicated situations where the entire field of view of an \textit{HST} observation may lie in a galaxy's disk, we use the upper main sequence stars as a proxy for other unwanted contaminants (e.g. red supergiants), and remove the regions with the highest relative number of upper-main sequence stars relative to other stars. This allows us to remove regions with the highest amounts of current star-formation as well as internal host extinction. In this manner, we are able to still select the most ideal subset of the observed field for our analysis. 

\subsection{Tip-Finding Methodology}

There are two main techniques used in the literature to measure the magnitude of the TRGB. Both involve constructing a luminosity function based on the color-magnitude diagram of stars. The abrupt onset of the helium flash at a predictable magnitude followed by the descent to the horizontal branch manifests itself as a discontinuity along the giant branch, and hence the observed luminosity function. The magnitude of this break is governed by the relative number of stars above and below the TRGB, i.e. the number of AGB and RGB stars. This exact ratio is dependent upon the star formation history of the galaxy.

The first technique for finding the TRGB involves applying an edge detector to the luminosity function. These edge detectors will elicit their maximum response at the sharpest discontinuities, including the TRGB. There are several options for edge detecting algorithms, but most applications in the literature use a form of the Sobel filter \citep{1993ApJ...417..553L,1995AJ....109.1645M,2019ApJ...875..136V}. In some cases, the luminosity function is first smoothed before the filter is applied \citep{2018ApJ...861..104H}.

The second commonly used technique is to fit the observations with a model luminosity function. This function often takes the form of a broken power-law, with the break signifying the magnitude of the TRGB. The physical basis for this is that the sharp transition of stars off the RGB imprints a discontinuity onto the luminosity function of the giant branch. The best-fit solution is found either by a maximum-likelihood method \citep{2002AJ....124..213M, 2016ApJ...826...21M}, or by a non-linear least squares method \citep{2019ApJ...872...80C}. A key benefit of this method is the ability to easily fold in results from the artificial star experiments, including bias, completeness, and photometric errors. Our own implementation of this technique is described in detail by \cite{2006AJ....132.2729M} and \cite{2014AJ....148....7W}. Previous works have found good general agreement between the two distinct techniques (see the comparisons in \citealt{2006AJ....132.2729M} and \citealt{2020ApJ...905..104K}).

The black line in the right panel of Figure \ref{NGC7640CMDS} shows our measurement of the TRGB for NGC 7640 ($m_{TRGB} = 25.74 \pm 0.05$). The break in the line shows the color range used for the fitting procedure. The color range, as in this case, is typically set at F606W-F814W = 0.5-2.0, though this may be shifted to the right in cases of significant reddening. The blue limit removes the main-sequence stars that lie at the same magnitudes, and the red limit removes high-metallicity RGB stars, for which the TRGB becomes intrinsically fainter (the situation reverses in the near-infrared, where the effects of line blanketing increase the intrinsic brightness).

\subsection{Foreground Reddening}

Our photometry, like many other forms of astronomical measurement, is impacted by interstellar reddening. We minimize extinction internal to the host galaxy by selecting regions in the halos, but this still leaves open the question of foreground extinction caused by looking through our own Milky Way galaxy. The vast majority of our targets lie at high galactic latitudes (b $> 15 \degree$), where the foreground extinction from the Milky Way is low. In these cases, we adopt a foreground reddening value from the dust maps of \cite{2011ApJ...737..103S}.

In the few cases where there is moderate to large foreground extinction at lower galactic latitudes, we take one of two approaches. The first, and more common method we use is to measure the displacement of the zero-age main sequence relative to a similar galaxy with a very well known foreground reddening value. The difference in color of the zero-age main sequences of two galaxies with similar metallicities provides a measurement of the relative reddening between the two targets. An example of this is highlighted in \cite{2017ApJ...835...78R}, where we use the very well known extinction to NGC~300 to measure the absolute extinction to ALFAZOA J1952+1428, a galaxy embedded within the Local Void.

In cases where the target is at an exceptionally low galactic latitude or the upper main sequence is not populated, we take another approach. Near the galactic midplane, the density of foreground stars in the observed color-magnitude diagram is often very high. In addition to their high density, they also occupy regions on the CMD that are not generally occupied by our extragalactic target stars. This allows us to use them to measure the foreground reddening. We can use TRILEGAL \citep{2005A&A...436..895G} to simulate the expected population of foreground stars in our observations, and then compare the offset from the observed foreground populations. This offset provides the amount of foreground extinction. For a recent example, see the galaxies studied in the near infrared presented in \cite{2019ApJ...880...52A}.

\subsection{Zero-point and Color Calibration}

Like any \textbf{other} standard candle, it is necessary to know the intrinsic brightness of the target object to determine its distance from us. The absolute magnitude of the TRGB thus requires a zero-point calibration. The zero-point we use has been determined from horizontal branch distances to nearby dwarf spheroidal galaxies \citep{2007ApJ...661..815R}. In addition to a zero-point calibration, there is also a color dependence to the TRGB. This dependence arises from the fact that the magnitude of the TRGB is sensitive to both metallicity, and to a lesser extent the age of the population. Both of these dependencies are projected onto the direct observable color \citep{2011AJ....141..106J,2019ApJ...880...63M}.  The \cite{2007ApJ...661..815R} paper provides both the zero-point and color calibrations in one relation. Independent relations are given for WFPC2 and ACS data, both in either F555W-F814W, or F606W-F814W colors. We use the median color of stars at the TRGB as the input color, with the errors determined via bootstrap resampling trials \citep{2014AJ....148....7W}. For the few optical WFC3/UVIS observations we have, we adopt WFC3 to ACS photometric transformations from \cite{2015ApJ...807..133J}. 

In addition to the optical bands, some of our data is taken in the near-infrared. This is typically the case when the target is at a low galactic latitude, such that the optical observations would be too severely affected by foreground extinction. Our near-infrared choice of calibration is presented by \cite{2014AJ....148....7W}, who use WFC3/IR data taken in F110W+F160W \citep{2012ApJS..198....6D}. The absolute calibration is then tied to the results from \cite{2007ApJ...661..815R}. The calibration is presented for both F110W and F160W bands. While observations in F160W are less sensitive to reddening, the absolute magnitude of the TRGB is nearly twice as sensitive to the color when compared to F110W.



\section{Data Access}

The Extragalactic Distance Database is hosted in Hawai'i by the Institute for Astronomy, and is available for public access at \url{https://edd.ifa.hawaii.edu}. The homepage of EDD gives descriptions for various catalogs, with newer highlights listed at the top. To see the list of catalogs, the end-user should click on the large green ``Next" button on the homepage. Here, the user is presented with a long list of catalogs that are hosted on EDD, otherwise known as the ``Select Table $\&$ Columns" page. The catalog of interest here is the CMDs/TRGB catalog, which is located under the banner of ``Stellar Distances". To see the entire catalog, the user should click the blank white square under the catalog title, and then hit the corresponding submit button below. Note that clicking on the catalog title itself will provide a description of each column given in the catalog's table. 

After clicking on the submit button, the user is taken to a ``Data Display" page. By default, 200 rows are shown, a number that can be changed by increasing the range of ``Requested Rows" on the top of the page, or by clicking ``All" instead of ``Submit" on the previous page. On this Data Display page, the user can choose to sort the catalog by the values of any of its columns, or proceed to download the entire table by following the download options on the bottom of the page. Clicking on a galaxy's name under the ``Name/CMD" column will bring up a new page with detailed information for that particular target. For example, clicking on NGC~7640 will bring up a new page with several pieces of information. The first is the row belonging to NGC~7640 from the larger catalog table that contains the relevant measurements for this target. Next is the color-magnitude diagram for the full field of the observation- clicking on this CMD will allow the user to retrieve the \textit{*.phot2} photometry file. This CMD is followed by the footprint of the field. In  cases where we restrict our observation to a limited region of the field, we show a ``Clipped CMD". The magnitude of the TRGB is marked with a black line- in cases with no Clipped CMD, we show this line on the full-field CMD. Lastly, we show a 600x600 color image of the \textit{HST} observation. A higher resolution version can be retrieved by clicking on this image.

There are many instances where galaxies have multiple datasets with photometry, and all of these remain archived on EDD. All of this information is repeated for each proposal/observation for the galaxy, except for the table row. In cases with multiple optical or multiple near-infrared observations, our preferred distance is the only one shown in the single table row at the top of the page. 

There is also the option to limit one's search to a single galaxy, instead of wading through the entire catalog. To do this, the user should simply navigate to the initial Select Table $\&$ Columns page, and type their galaxy of choice into the box labeled ``OPTIONAL: Enter Galaxy Name". Clicking the button next to this will then display only the catalogs which have information on the selected galaxy. Our backend (which uses NED\footnote{\url{ned.ipac.caltech.edu}} to resolve the name) will handle most well-known galaxy catalog names (e.g. Messier, NGC, ESO, etc). However, to guarantee a match, we recommend that users use a galaxy's Principal Galaxy Catalog (PGC) number, which can be easily be found by visiting the HYPERLEDA database\footnote{\url{http://leda.univ-lyon1.fr/}} \citep{2014A&A...570A..13M}. A key benefit of EDD is that the individual catalogs can all be linked together by this PGC number. One can easily obtain information from multiple catalogs for a single target by clicking the check-boxes on any number of catalogs for a galaxy, and then submitting their request.

We note that the CMDs/TRGB catalog is updated on a regular basis, and that users should check EDD to see if previous literature distances have been updated in our catalog. Similarly, we encourage users to contact us if there is any discrepancy noted within our database, or with any requests for analysis of particular datasets.

In addition to the regular availability through EDD, we are currently engaged in efforts with MAST to provide the CMDs/TRGB catalog through the High Level Science Product interface (HLSP\footnote{\url{https://archive.stsci.edu/hlsp/}}) in the coming months. 



\section{Science Highlights}

\begin{figure*}
\centering
\epsscale{1.1}
\plotone{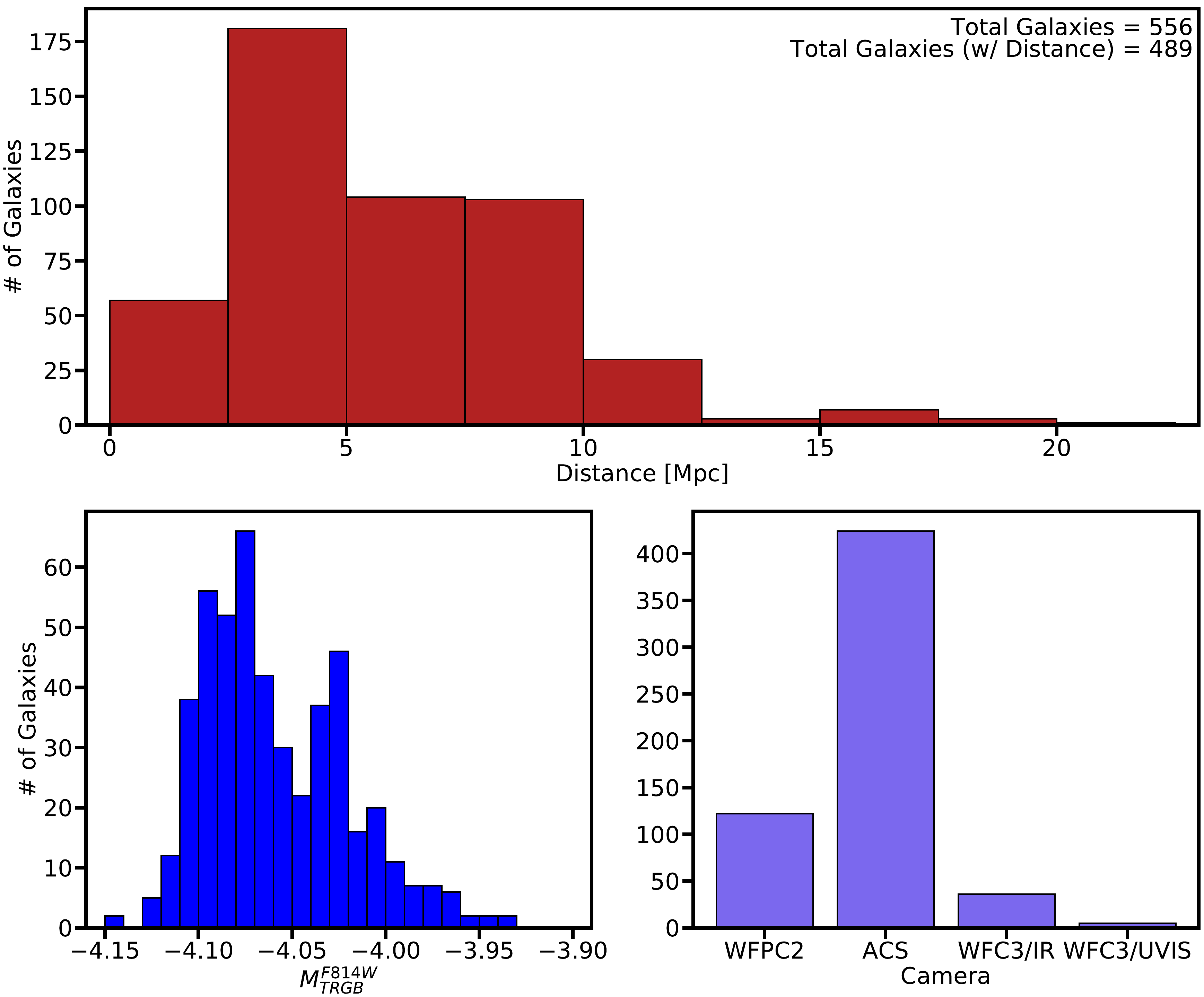}
\caption{\textbf{Top)} Histogram of TRGB distances for galaxies within the CMDs/TRGB catalog. 489/556 of the galaxies have a TRGB distance measurement-- for galaxies with both optical (F814W) and near-infrared (F110+F160W) measurements, we simply display the median. \textbf{Bottom Left)} A histogram of the absolute magnitude of the TRGB determined with the \cite{2007ApJ...661..815R} for galaxies with optical (F814W) TRGB measurements. A majority of the targets within our catalog (and within the Local Volume) are metal-poor dwarf galaxies, for which $M^{F814W}_{TRGB}$ is preferentially brighter. \textbf{Bottom Right)} A bar plot showing the distribution of \textit{HST} imaging in our sample. In cases with multiple optical observations (i.e. WFPC2 and ACS), we only show the camera which is used for the final determination of the TRGB. The sum total of the bar plot (587) exceeds the total number of galaxies (556) as there are targets which have data available in both optical and near-infrared wavelengths.}
\label{barPlot}
\end{figure*}

The \textit{HST} archives contain a wealth of data for the science of resolved stellar populations. We aim to provide photometry and a TRGB distance for every galaxy with appropriate data. A summary figure showing the current state of our catalog can be seen in Figure \ref{barPlot}. The top panel shows a histogram of the distribution of galaxies for which we have been able to successfully derive TRGB distances (489/556 of the galaxies in our catalog). For galaxies with multiple distances (derived separately in F814W, F110W, and F160W), we simply display the median. While the majority of our targets are located within 10~Mpc, deep data from several programs allow us to measure distances for galaxies out to just past 20~Mpc. For the remaining galaxies (67/556), we are not able to derive a TRGB distance with the data available. In these cases, we still provide color-magnitude diagrams and the underlying photometry via our catalog.

The bottom-left panel shows the distribution of derived absolute TRGB magnitudes for galaxies with optical (F814W) data. While several others find an absolute TRGB zero-point calibration that is between $M^{F814W}_{TRGB} \sim-4.05$ to $-3.97$ (see $\S$7.3 for an in-depth discussion), our use of the \cite{2007ApJ...661..815R} zero-point and color calibration provides us with a median of $M^{F814W}_{TRGB}=-4.07$. This is to be expected, as most of the galaxies in our catalog (and in the Local Volume) are metal-poor, dwarf galaxies. Lastly, the bottom-right panel shows the number of galaxies which have data that are used for purposes of measuring the TRGB from each of four individual detectors on \textit{HST}.  For cases with multiple optical observations (eg. WFPC2 and ACS), we only tabulate the camera which is used for the final distance determination in the catalog table. In all cases, the CMDs/TRGB catalog table displays the best available distance in each of optical (F814W) or near-infrared (F110W+F160W) bands.

It is always useful to have an independent check on our catalog values, as agreements between separate groups improve confidence in the TRGB as a high-precision distance measurement technique. The set of 8 TRGB distances to nearby spiral galaxies presented in \cite{2017AJ....154...51M} provide a good basis for such a comparison. As they use the same absolute calibration in their work as we do in our catalog \citep{2007ApJ...661..815R}, we can compare the final values of distance modulus directly to one another. Using the present CMDs/TRGB catalog values, we find a very small average (both mean and median) offset of  $\mu_{EDD}-\mu_{McQuinn}$ = $-$0.01 mag, with a scatter of 0.08 mag. The two independent determinations are in excellent agreement.

In a few instances, there are cases where literature TRGB measurements may be significantly different from the ones reported in our catalog. One such example is of the dwarf galaxy Coma~P.  \cite{2018AJ....155...65B} first reported a preliminary TRGB measurement to Coma~P of 5.50 $\pm$ 0.28~Mpc, which was later published by \cite{2019AJ....157...76B}. We reported the CMDs/TRGB catalog measurement of this galaxy in \cite{2018ApJ...861L...6A}, where we found D = 10.9 $\pm$ 1.0~Mpc. The factor of $\sim$2 difference in the distance is substantial, and is the result of uncertainty in what constitutes the RGB. We consider the discontinuity at $m_{TRGB}\sim$24.6 mag to indicate the onset of AGB stars, whereas \cite{2019AJ....157...76B} take it to be the TRGB. This confusion typically results from data that is not sufficiently deep (as is the case here), or with data in highly crowded regions with substantial populations of younger stars \citep{2014AJ....148....7W,2018AstBu..73..279T,2019ApJ...872L...4A}. A third independent analysis of the same \textit{HST} data by \cite{2020AstBu..75..103T} finds a distance similar to ours (D = 12.7 $\pm$ 0.9~Mpc).  In the case of Coma~P, a definitive measurement of the TRGB may only be possible with deeper \textit{HST} data. The large disagreement in the available TRGB distances for Coma~P is a rare case, and there is generally very good agreement between various TRGB measurements for galaxies, as shown by the comparison between our work and that of \cite{2017AJ....154...51M}.

In the rest of this section, we describe science cases that showcase the science possible with the information available in our catalog.

\subsection{Enabling a Broad Range of Science}

The data within the CMDs/TRGB catalog has enabled a large breadth of science by our collaboration and others. The distances obtained from the TRGB serve as important calibrators for methods such as the Tully-Fisher relation \citep{2020ApJ...896....3K}, and also serve as an important cross-check to the Cepheid distance ladder \citep{2016ApJ...826...56R}. These TRGB distances serve as part of an important foundation for the broader Cosmicflows program \citep{2008ApJ...676..184T, 2016AJ....152...50T}, which has been instrumental in determining the extragalactic distance scale, studying large-scale structure, and determining the value of the Hubble Constant. Most recently, preliminary results from Cosmicflows-4 indicate a value of $H_{0}$ $\sim$75 km/s/Mpc \citep{2020ApJ...896....3K,2020ApJ...902..145K}. In a future work (Anand et al., in prep), we will also provide a detailed comparison of our TRGB results to those obtained by the Carnegie-Chicago Hubble Program \citep{2019ApJ...882...34F}.

The knowledge of so many highly accurate ($\sim$5$\%$) distances has allowed us to reconstruct the large-scale structure of our local Universe. Combining these distances with velocity information, we are able to numerically reconstruct the orbital histories of galaxies from a numerical action methods (NAM) model \citep{2017ApJ...850..207S}. NAM gives a plausible solution for the evolution of galaxy orbits from $z$=4 to the present day, providing key constraints on past galaxy and group interactions and the dynamical evolution of large-scale structure such as the Local Sheet \citep{2019ApJ...880...52A}. Along with such efforts, it has also become clear the large extent to which our location nearby the Local Void has influenced the dynamical history of the Milky Way and many of its nearby counterparts \citep{2008ApJ...676..184T, 2019ApJ...880...24T}.

The $\sim$5$\%$ accuracy of TRGB distances at the one orbit limit of 10~Mpc translates to 500~kpc in physical terms. This scale is comparable to the virial radius of large galaxies. This fact allows us to unequivocally associate galaxies into distinct groups and study the group properties of different systems within the Local Volume \citep{2005AJ....129..178K,2014AJ....147...13K}. As the second turnaround radius of the Virgo Cluster is $\sim$9~Mpc from us, we have also used some of the more distant objects in our sample to map the infall of galaxies onto the cluster \citep{2018ApJ...858...62K}. As an additional bonus, there have been instances where new dwarf galaxies were first discovered as part of our TRGB program \citep{2018MNRAS.474.3221M}.

The record of stellar content contained within CMDs of the target galaxies is useful not only for measurements of the TRGB; they also allow us to quantify the evolutionary histories of stars within these galaxies. Ongoing star formation is manifested in the bright upper-main sequence \citep{2012MNRAS.423..294M, 2017MNRAS.464.2281M}, whereas the red giant branch and fainter horizontal branch provide constraints on the evolution of the oldest stars \citep{2011AJ....141..106J}. In the near-future, we intend to develop upon our existing tools \citep{2003ApJ...589L..85R, 2011AJ....141..106J} and present star-formation histories for many of the galaxies available within our catalog.

\begin{figure*}
\centering
\epsscale{1.}
\plotone{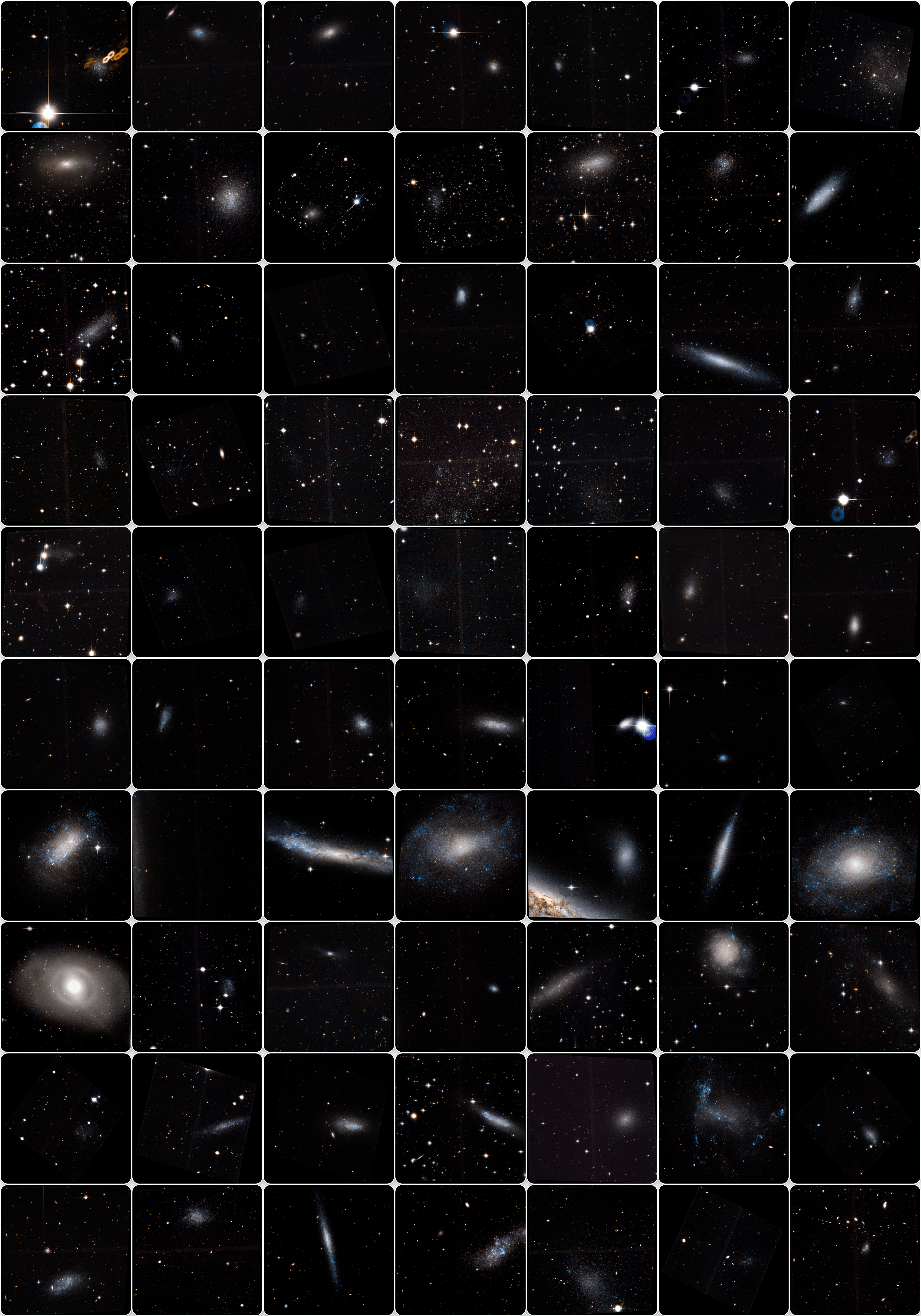}
\caption{A collage of ACS color (F606W+F814W) images for the sample of galaxies observed as part of the Every Known Nearby Galaxy survey (SNAP-15922, PI R. Brent Tully). While most of the observed galaxies are small dwarfs, we see that this program also provides imaging and distances to nearby large spirals. High resolution versions of each of these images are available on EDD.}
\label{collage}
\end{figure*}

\subsection{Every Known Nearby Galaxy}
\textit{HST} has the unique capability to determine a TRGB distance to essentially \emph{any} galaxy within 10~Mpc with just a single orbit of observations. The larger James Webb Space Telescope (\textit{JWST}) will be disfavored for observations of nearby galaxies due to inefficient setup/exposure ratios. It would thus be a fitting legacy for \textit{HST} to have observed \emph{every} known galaxy thought to lie within 10~Mpc, not only to derive an accurate distance but also to have a record of its stellar content. The distances provide key constraints on group dynamics, local large-scale structure and flows, and information on satellite luminosity functions, among other topics of interest. The stellar content on the other hand provides important information on the star formation histories of these galaxies, providing vital insights into the evolutionary histories of nearby galaxies. To this end, our collaboration has been applying for and competitively awarded time on \textit{HST} through multiple General Observer (GO) and snapshot (SNAP) proposals which have provided observations for hundreds of nearby galaxies.

Most recently, we were awarded a Cycle 27 SNAP proposal (SNAP-15922) to observe a selection of the remaining 153 \emph{known} galaxies thought to lie within 10~Mpc. This aptly named ``Every Known Nearby Galaxy" survey is nearing completion, and has so far provided observations for 70 known nearby galaxies. A montage of color images from this program can be seen in Figure \ref{collage}. While most of the galaxies shown are dwarfs, a few large galaxies (e.g. NGC 4242, NGC 5585) are also seen. Key results from this program are being published separately \citep{2020A&A...643A.124K, 2020A&A...638A.111K, 2021arXiv210211354K}.  Along with the release of this paper, we have made our photometry and TRGB determinations for all of these targets publically available on the CMDs/TRGB catalog.

\subsection{Serendipitous Observations of a Local Volume Dwarf}

\begin{figure*}
\centering
\epsscale{1.1}
\plotone{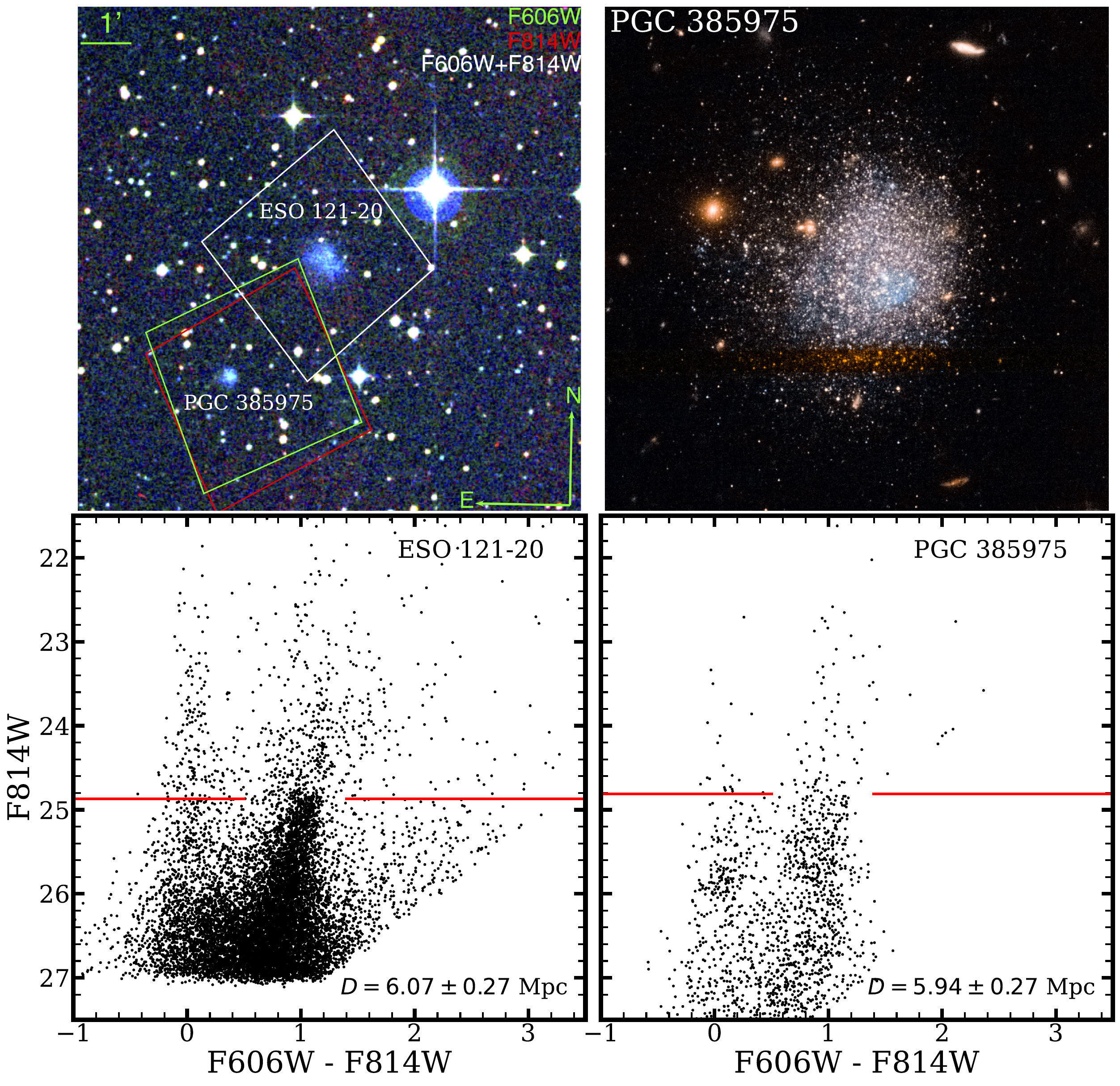}
\caption{\textbf{Top Left)} Archival HST pointings of ESO~121-20 and PGC~385975 overlaid on ground-based DSS imaging. \textbf{Top Right)} ACS color (F606W+F814W) image of the serendipitous observations of PGC~385975. \textbf{Bottom Left)} Color-magnitude diagram and TRGB determination for ESO~121-20. Besides the red giant branch, we also see a large amount of asymptotic giant branch and young upper-main sequence stars. \textbf{Bottom Right)} Color-magnitude diagram and TRGB determination for PGC~385975. While sparser than its neighboring counterpart, we are still able to determine a precise distance to the target and confirm that the two galaxies are physically associated.}
\label{ledaQuadPlot}
\end{figure*}

Given the broad scope of resolved stellar population science, there are many observations of nearby galaxies not taken for the explicit purpose of deriving a TRGB distance. In some instances, the observations of these galaxies themselves are serendipitous! Here we highlight one such case for PGC~385975, a Local Volume dwarf for which data was captured while observing a background galaxy cluster.

SPT-CLJ0615-5746 is a z=0.972 \citep{2016A&A...594A..27P} galaxy cluster, observed as part of the Reionization of Lensing Clusters Survey (RELICS) collaboration \citep{2019arXiv190302002C}. Fortuitously, the dwarf galaxy PGC 385975 lies nearby and in the foreground of this cluster, and two orbits of optical observations (split between F606W $\&$ F814W) covering this dwarf were taken as part of the programs GO-12477 (PI F. High) $\&$ GO-12757 (PI P. Mazzotta).

PGC~385975 was first discovered during the course of HI observations of a nearby larger dwarf galaxy, ESO~121-20 \citep{2006AJ....131.2056W}. These observations, taken with the Australia Telescope Compact Array (ATCA), revealed this companion to be separated from ESO~121-20 by 3$'$ on the sky, and with a systemic velocity difference of only 29 $\pm$ 6 km/s. \cite{2014AJ....147...13K} maintain a catalog of Local Volume galaxies and their group membership status based on a tidal index ($\Theta_{1}$) calculated using K-band luminosities and physical separations of galaxies from one another. $\Theta_{1} > 0$ indicates two galaxies are physically bound. They list the main disturber of PGC~385975 as ESO~121-20 and assign it a tidal index of $\Theta_{1} = 0.6$, indicating the two are likely bound, with the assumption that the two lie at the same distance.

To confirm the physical association of these two dwarfs, we reduced and analyzed the \textit{HST} data with the serendipitous observations of PGC~385975. The top left panel of Figure \ref{ledaQuadPlot} shows a DSS color image of the two galaxies, centered on ESO~121-20. Overlayed are footprints of ACS observations of both of the targets. ESO~121-20 was observed with ACS as part of SNAP-9771 (PI I.Karachentsev), and has a distance of 6.07 $\pm$ 0.27~Mpc from our previous measurement of its TRGB (bottom left panel). The observations of PGC~385975 were focused on a background cluster (not visible in this DSS image), and do not perfectly overlap. Nonetheless, we are able to construct the CMD for this galaxy (color image in the top-right panel), from which we find $m_{TRGB}$ = 24.81 $\pm$ 0.06 and determine a distance of ${\rm D}=5.94\pm0.27$~Mpc (bottom right panel). The CMD for ESO 121-20 shows a prominent upper-main sequence, indicating that star-formation is still ongoing. PGC~385975 has fewer upper-main sequence stars, and consists of mainly an older stellar population. There are two interesting features of note in the CMD for PGC~385975. One is that despite the fact that ESO~121-20 ($M_{B}$ = -13.82\footnote{Magnitudes rescaled from \url{https://relay.sao.ru/lv/lvgdb/}}) is significantly brighter than PGC~385975 ($M_{B}$ = -12.03), it is PGC~385975 which has the broader red giant branch. The second is that there is a visible discontinuity in both the young and old stellar populations at around m$\sim$26. One explanation that resolves both of these peculiarities could be a recent burst of star formation which began between 100-200~Myr ago.

With our TRGB distance to PGC~385975, we confirm that ESO~121-20 and PGC~385975 are members of the same dwarf association, of which they are the only known constituents. While this situation was previously considered likely, there have been cases in the literature where two galaxies with similar velocities and small angular separations were assumed to be part of the same group, but later found to be separated by several Mpc (e.g. NGC~4618 and NGC~4625 \citep{2017AJ....154...51M, 2018ApJ...858...62K}). At the projected separation of 3$'$, the physical distance between these two dwarfs is $\sim$5 kpc, though this is a lower limit as it assumes the same physical distance to the two. This pair of dwarfs is reminiscent of another pair of binary dwarf galaxies, LV~J1157+5638 and its satellite which are located a distance of 9~Mpc from us and physically separated by 4~kpc from each other \citep{2018MNRAS.474.3221M}.

Along with their HI observations, \citep{2006AJ....131.2056W} also performed follow up optical observations with the 2.3m Australian National University telescope. They found no evidence of either an HI or stellar bridge connecting these two dwarfs. The \textit{HST} field of view for PGC~385975 extends all the way to the southeastern edge of ESO~121-20 (see top left panel of Figure \ref{ledaQuadPlot}), allowing us to look more carefully for hints of previous interactions. With the much fainter limiting magnitude and higher resolution of our data, we also find no evidence of any stellar stream outside of the main bodies of the galaxies, suggesting that any previous interactions between the two has been limited. 

\begin{figure*}[]
\centering
\epsscale{1.15}
\plotone{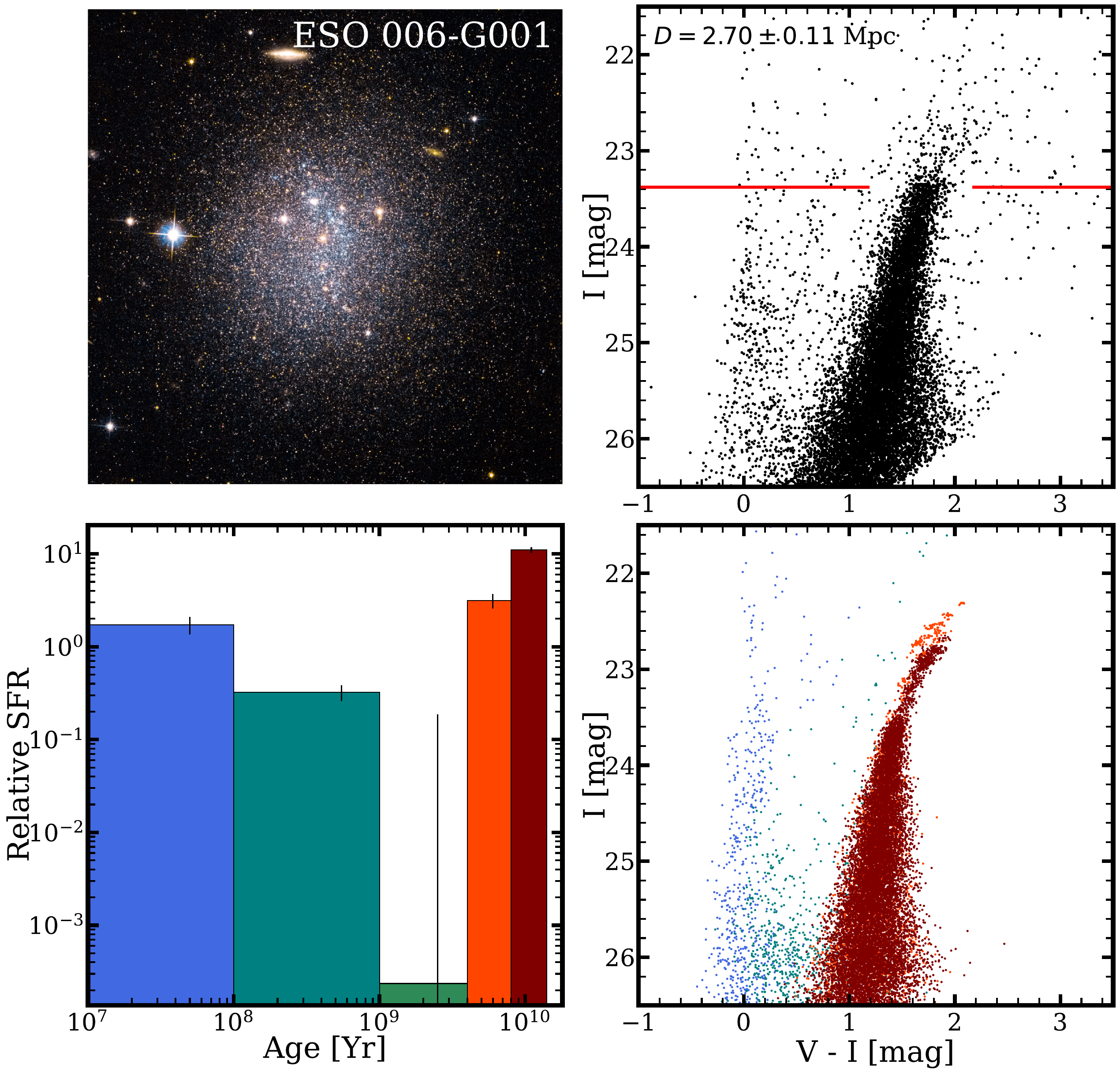}
\caption{\textbf{Top Left)} An ACS color (F606W+F814W) image of the nearby dwarf galaxy ESO~006-G001. Visible are both young (blue) and old (red) stellar populations. \textbf{Top Right)} The color-magnitude diagram and derived TRGB measurement for the galaxy. The CMD shows a prominent red giant branch, as well as a modest amount of upper-main sequence stars which indicate the presence of ongoing star formation. \textbf{Bottom Left)} Our derived star formation history, which is normalized to a rate of 1000 $M_{\odot}$ $Myr^{-1}$. The left hand side of the age range for the most recent bin is limited to $10^{7}$ years for the purposes of plotting, but our star-formation history analysis extends to the present day. \textbf{Bottom Right)} The simulated CMD for ESO~006-G001 used to derive the galaxy's star-formation history.}
\label{SFH}
\end{figure*}

\subsection{The Star Formation History of a Transition Dwarf Galaxy}

As mentioned earlier, the color-magnitude diagrams available in our catalog provide a record of the stellar content of each of the target galaxies. Encoded within these CMDs are the star-formation histories, which can be quantitatively measured by creating synthetic stellar populations and attempting to best fit the observed CMD \citep{2003AJ....126..187D, 2011AJ....141..106J, 2014ApJ...789..147W, 2015ApJ...802...66M}. In this subsection, we derive the star formation history of ESO~006-G001.

ESO~006-G001 (PGC~23344) is a nearby dwarf galaxy with a radial velocity of $v_{hel}$ = 319$\pm$58 km/s (R. Kraan-Korteweg and T. Lambert, priv. communication). \cite{2002A&A...388...29P} obtained B and R photometry of the galaxy and found that the outer regions were reminiscent of a dwarf elliptical, whereas the center contained bright knots. Based on this, they postulated that ESO~006-G001 is likely a transition dwarf galaxy. We obtained observations (760s each in F606W and F814W) of ESO~006-G001 as part of SNAP-15922 (PI R. Tully). The top-left panel of Figure \ref{SFH} shows a color image limited to the region of the dwarf galaxy. While the majority of stars appear red (and hence old), there are a number of stars which are blue and are indicative of ongoing star-formation. The CMD for the galaxy (top-right panel of Figure \ref{SFH}) also shows a distinct red giant branch, with a sharp discontinuity at the location of the TRGB. We measure $m_{TRGB}$ = 23.37  $\pm$ 0.02, from which we find a distance of D = 2.70$\pm$ 0.11~Mpc. 

As our data for this galaxy does not extend down to the horizontal branch, our age resolution for it's derived star formation history is limited for all but the youngest stars. In general,  we follow the methods detailed in \cite{2011AJ....141..106J}. Briefly, we use the Padova isochrones \citep{1994A&AS..106..275B, 2000A&AS..141..371G} to simulate a large range of stellar populations, with varying ages and metallicities. These simulated populations are convolved with the photometric errors and completeness that are determined by our artificial stars experiments. Initial constraints on age and metallicity are determined by using the observed features (e.g. width of the red giant branch, relative number of main sequence stars) in the \textit{HST} color-magnitude diagram. The observed CMD is split into bins based on the separate stellar sequences (see Figure 5 in \citealt{2011AJ....141..106J}), and the best-fit star formation history is determined via a multi-dimensional Amoeba-style minimization algorithm.

Our results are shown in the bottom panels of Figure \ref{SFH}. The bottom-left panel shows the derived star-formation rates, which are normalized to a rate of 1000 $M_{\odot}$ $Myr^{-1}$. The bottom-right panel shows the best-fit CMD for the target. We find that ESO~006-G001 had it's highest rate of star formation at ancient times, and that this rate underwent a decline till ~$\sim$1 Gyr ago (though there is great uncertainty as to the magnitude of the decline). We also find that the rate of star formation has been elevated for the last ~$\sim$1 Gyr, as evidenced by the current star-forming regions within the galaxy. We reserve an in-depth discussion of the star formation history of this galaxy as well as it's motion to a separate, future work.



\section{Future Outlook}

\subsection{Archival Material}

Our intent is to have the CMDs/TRGB catalog continuously updated, by continuing to analyze all galaxies that have the requisite data in the \textit{HST} archives. As of the writing of this paper, the CMDs/TRGB catalog is essentially up-to-date, with only a handful of difficult cases left to analyze. We continue to keep a close eye on the archive for new material that is suitable for the purposes of determining TRGB distances. We encourage the community to reach out to us with data that we may have overlooked.

\subsection{The State of TRGB Calibrations}

Over a decade has passed since the \cite{2007ApJ...661..815R} paper, and there have been several other versions of an optical calibration since then. We use the this subsection to discuss these, as well as the future of the TRGB calibration. 
 
As part of a series of papers \citep{2015ApJ...807..133J, 2017ApJ...836...74J} that investigated the use of the TRGB in determining the value of the Hubble Constant,  \cite{2017ApJ...835...28J} present a new calibration of the absolute magnitude of the F814W TRGB which is quadratic in nature. While this functional form appears to be more in line with what is predicted in theoretical models (e.g. \citealt{2004ApJ...612..168P}), we argue that the quadratic shape of their calibration is influenced by their extremely high metallicity measurements in large spiral galaxies. In their Figure 5, several data points extend out to ${(F606W-F814W)}_{0}=3.5$. In practice, our catalog measurements rarely probe beyond a red limit of ${(F606W-F814W)}_{0}=2.0$ (with a median closer to $\sim$1.2), a regime where the \cite{2017ApJ...835...28J} data points exhibit a trend that is much closer to linear in nature. There is also a small but noticeable difference between these two calibrations in the blue color range corresponding to metal-poor systems. The \cite{2007ApJ...661..815R} calibration indicates a steeper slope for $M_{TRGB}$ compared to the more constant value indicated by \cite{2017ApJ...836...74J}. For the most metal-poor systems found in our catalog (F606W-F814W$\sim$0.9), the differences in the value of $M_{TRGB}$ may be as large as $\sim$0.1~mag, or $\sim$5$\%$ in distance. Such discrepancies only apply to the most metal-poor systems, as the two calibrations give nearly identical values of $M_{TRGB}$ for the average galaxies in our catalog.

More recently, there has been a trend towards a single absolute calibration. The underlying assumption here is that if one measures the TRGB in relatively metal-poor regions (i.e. far out into the halos of spiral galaxies), the absolute magnitude should be essentially constant. As part of the Carnegie-Chicago Hubble Program, \cite{2019ApJ...882...34F} determine a value of $M_{F814W}^{TRGB} = -4.049$ $\pm$ 0.022 (stat) $\pm$ 0.039 (sys) from a measurement of the TRGB in the LMC, which has a highly precise geometric distance from measurements of detached eclipsing binaries \citep{2019Natur.567..200P}. In response to this calibration, \cite{2019ApJ...886...61Y} argue that the extinction in the LMC was overestimated, and find a value of $M_{F814W}^{TRGB} = -3.97$ $\pm$ 0.046. \cite{2020ApJ...891...57F} present an analysis of the total line-of-sight extinction to TRGB stars in the LMC and find $M_{F814W}^{TRGB} = -4.054$ $\pm$ 0.022 (stat) $\pm$ 0.39 (sys). \cite{2021ApJ...908L...5S} use recently released maps of LMC extinction \citep{2021ApJS..252...23S} and find $M_{F814W}^{TRGB} = -3.96$. Given the dusty nature of the LMC, this is clearly a difficult measurement, and the matter is still being played out in the literature. 

Besides the issues of extinction, the bulk of the photometry that has been used to measure the TRGB in the LMC is ground-based. This has been necessary because to gather enough TRGB stars in the LMC, one needs to cover a large area on the sky, making direct measurements with \textit{HST} prohibitively expensive thus far. This issue introduces two sources of systematic errors. The first involves the conversion of filter systems (from ground-based to the \textit{HST} flight system), while the second arises from blending in the ground-based photometry due to limits in imaging resolution. While both of these issues have been considered by the papers mentioned above, a more precise calibration of the TRGB in the LMC will be difficult to obtain given these circumstances. 

Other efforts for a singular zero-point calibration of the TRGB have targeted NGC~4258, which is host to a water megamaser. This allows for the determination of a highly precise geometrical distance of D = 7.576 $\pm$ 0.082 (stat) $\pm$ 0.076 (sys) Mpc \citep{2019ApJ...886L..27R}. \cite{2019ApJ...886L..27R} themselves average two previous TRGB measurements by \cite{2006ApJ...652.1133M} and \cite{2017ApJ...835...28J} to establish $M_{F814W}^{TRGB} = -4.01$ $\pm$ 0.04. On the other hand, \cite{2021ApJ...906..125J} perform an analysis of a large set of archival \textit{HST} data in the outer halo of NGC~4258 from which they determine $M_{F814W}^{TRGB} = -4.050$ $\pm$ 0.028 (stat) $\pm$ 0.048 (sys). Our own catalog distance to NGC~4258 is $\mu = 29.42 \pm 0.04$ (stat) with an adopted value of $M_{F814W}^{TRGB} = -4.01$ determined via the \cite{2007ApJ...661..815R} calibration. This is in good agreement with the megamaser distance modulus of $\mu = 29.397 \pm 0.032$ \citep{2019ApJ...886L..27R}, which when combined with our measurement would shift the \cite{2019ApJ...886L..27R} zero-point to $M_{F814W}^{TRGB} = -3.99 \pm 0.04$.  

Besides the LMC and NGC~4258, where else can we turn to for an absolute calibration for the TRGB? Initial results from the Gaia spacecraft show great promise \citep{2016A&A...595A...1G,2018A&A...616A...2L}. Efforts are already underway \citep{2019PASA...36....1M}, including some that present an initial calibration \citep{2020arXiv201209701C,2021ApJ...908L...5S}. \cite{2020arXiv201209701C} use Gaia DR2 data to determine cluster memberships for stars in 46 Milky Way globular clusters, and then measure the TRGB from a composite CMD to find $M_{I}^{TRGB} = -4.056$ $\pm$ 0.02 (stat) $\pm$ 0.10 (sys). \cite{2021ApJ...908L...5S} instead measure the parallax to the Milky Way globular cluster $\omega$ Cen, and use an existing TRGB measurement \citep{2001ApJ...556..635B} to determine an absolute magnitude of the TRGB of $M_{I}^{TRGB} = -3.97$ $\pm$ 0.06. The differences between these two studies can be explained by the difference in values of $\mu$ for $\omega$ Cen. Whereas \cite{2021ApJ...908L...5S} measure a value of D = 5.24 $\pm$ 0.11~kpc directly from the Gaia EDR3 data, \cite{2020arXiv201209701C} use a value of D = 5.44 $\pm$ 0.28~kpc adapted from observations of detached eclipsing binaries \citep{2001AJ....121.3089T}. The distance to $\omega$ Cen is a value that will certainly be better known as the Gaia mission continues and the existing systematic issues are better understood.  Regardless, for our catalog we require a calibration that includes a color-dependent term. The Local Volume contains galaxies which span a broad range of metallicities and ages, and there are many archival pointings we use which are not the most optimal (e.g. pointings in high metallicity spiral disks). 

For now, we continue to use the color and absolute calibration of \cite{2007ApJ...661..815R}, and intend to revisit this topic in the near-future with a multitude of techniques. Since our catalog provides all the necessary intermediate measurements (e.g. $m_{TRGB}$, median color of the measured TRGB, adopted E(B-V), etc.), our distances can easily be swapped to a different calibration of $M_{TRGB}$ in the future.

\subsection{JWST $\&$ Roman Space Telescope}

While \textit{HST} may very well last into the 2030's, we will soon run into the cases where it would be more efficient to measure TRGB distances with future facilities such as \textit{JWST} and the Nancy Grace Roman Space Telescope (\textit{Roman}). JWST is scheduled to launch at the end of 2021 and will provide a significant advantage over \textit{HST} in many respects. Its larger aperture and near-infrared capabilities will make it possible for TRGB distances to be easily measured out to 30--40 Mpc with modest exposure times. It will be possible to measure accurate distances to galaxies within a significantly larger volume than is currently feasible. The main difficulty will be accurately calibrating the absolute magnitude of the TRGB in the near-infrared bands. As shown in many previous works \citep{2012ApJS..198....6D, 2014AJ....148....7W, 2019ApJ...880...63M, 2020ApJ...898...57D}, the absolute magnitude in the near-infrared can vary greatly as a function of color (a projection of metallicity and age effects). An in-depth calibration effort will be required to be able to extract the optimal science out of \textit{JWST}.

The Nancy Grace Roman Space Telescope will have an aperture identical to that of \textit{HST} but will offer a significantly larger field of view (0.28 $\mathrm{deg^{2}}$) associated with its main Wide-Field Instrument (WFI).  With the WFI, it will be feasible to both simultaneously discover and characterize satellite systems around nearby large galaxies. In addition to the boon for studying satellite systems, we will also be able to obtain coverage of entire spiral disks in a single pointing. This will allow us to precisely characterize the structure of the TRGB as a function of metallicity and galactocentric radius, important for the high precision needed to accurately determine the value of the Hubble Constant.


\section{Summary}

We present updated details on the CMDs/TRGB catalog that is available on the Extragalactic Distance Database. The catalog has now more than doubled in size to contain 556 galaxies, of which 489 targets have a measure of distance from the TRGB. Our uniform reduction and analysis procedures allow for a direct comparison between all of the galaxies in our sample, and we make all of our photometry publicly available along with color-magnitude diagrams, footprints, and color images.

The information within the CMDs/TRGB catalog has provided the foundation for transformative science within the Local Volume, and will continue to expand as more observations become available with \textit{HST}. In the near-term, we look forward to revisiting the issue of both the zero-point and color calibration of the TRGB, especially given its recent importance in potentially resolving the tension in the value of the Hubble Constant. Given the combined capabilities of \textit{Roman} and \textit{JWST} in the near-infrared, we look forward to a future for the TRGB that is bright.


\begin{acknowledgements}

This research is supported by an award from the Space Telescope Science Institute in support of program SNAP-15922. P.F.W. acknowledges the support of the fellowship from the East Asian Core Observatories Association. We acknowledge the usage of the valuable HyperLeda database (http://leda.univ-lyon1.fr). This research has made use of the NASA/IPAC Extragalactic Database (NED), which is operated by the Jet Propulsion Laboratory, California Institute of Technology, under contract with the National Aeronautics and Space Administration. We thank the anonymous referee for useful comments which improved this manuscript.  G.A. thanks Zach Claytor, Ryan Dungee, and Adam Riess for useful discussions, and Trystan Lambert and Ren\'{e}e Kraan-Korteweg for the redshift measurement of ESO~006-G001.

\end{acknowledgements}

\software{APLpy \citep{2012ascl.soft08017R}, DOLPHOT \citep{2000PASP..112.1383D,2016ascl.soft08013D},  DrizzlePac \citep{2015ASPC..495..281A}, Montage \citep{2010ascl.soft10036J, 2017ASPC..512...81B}}

\bibliography{paper}
\bibliographystyle{aasjournal}

\end{document}